\documentclass[a4paper,11pt]{article}
\pdfoutput=1 

\usepackage{jheppub}

\makeatletter
\def\@fpheader{\relax}
\makeatother

\title{\boldmath Index theory and dynamical symmetry enhancement of M-horizons}

\author[a]{J. Gutowski}
\author[b]{and G. Papadopoulos}
\affiliation[a]{Department of Mathematics \\
University of Surrey \\
Guildford, GU2 7XH, UK}
\affiliation[b]{Department of Mathematics\\
King's College London\\
Strand\\
London WC2R 2LS, UK}

\emailAdd{j.gutowski@surrey.ac.uk}
\emailAdd{george.papadopoulos@kcl.ac.uk}

\abstract{
We show that near-horizon geometries of 11-dimensional supergravity
preserve  an even number of supersymmetries. The proof
follows from Lichnerowicz type theorems for two horizon Dirac operators, the field equations and Bianchi identities,  and  the   vanishing
of the index of a Dirac operator on the 9-dimensional horizon sections.
As a consequence of this, we also prove that all M-horizons with non-vanishing fluxes admit a $\mathfrak{sl}(2, \mathbb{R})$ subalgebra of symmetries.}

\begin{document}
\maketitle
\flushbottom

%%%%%%%%%%%%%%%%%%%%%%%%%%%%%%%%%%%%%%%%%%%%%%%%%%%%%%%%%%%%%%%%%%%%%%%%%%

%%%%%%%%%%%%%%%%%%%%%%%%%%%%%%%%%%%%%%%%%%%%%%%%%%%%%%%%%%%%%%%%%%%%%%%%%%
\section{Introduction}

It is well known that
 the near horizon geometries of supersymmetric black holes and branes exhibit supersymmetry enhancement. There are many examples of this,
which include the RN black hole as well as the D3-, M2- and M5-branes, see eg \cite{gibbons}. In all these cases, while the black hole and brane configuration preserve half the supersymmetry,
the near horizon geometries are maximally supersymmetric.   If supersymmetry enhancement near the
horizons is universal\footnote{Throughout this paper we consider supergravity theories without higher curvature corrections.}, all near horizon geometries must preserve at least two supersymmetries.

The near horizon geometries as expressed in the  Gaussian null coordinates of \cite{isen, gnull} exhibit two continuous symmetries. For example, the M-horizon fields
 in (\ref{mhm}) are invariant under the continuous symmetries generated by the vector fields $\partial_u$ and $u\partial_u-r\partial_r$. However in all known examples, the near horizon geometries
 exhibit additional symmetries, and the two above symmetries enhance at least to $\mathfrak{sl}(2, \mathbb{R})$.  The presence of these additional symmetries cannot be explained from the kinematics of the system. Therefore, it is a consequence of the dynamics and  arises after
 implementing the field equations.

To our knowledge despite the overwhelming evidence there is for symmetry enhancement for near horizon geometries,   there is no general proof which demonstrates this. For example, it is not known  whether the near horizon geometries which preserve one supersymmetry automatically  preserve two or more. The same applies for the additional bosonic symmetries, and in particular $\mathfrak{sl}(2, \mathbb{R})$,  of near horizon geometries.

Symmetry enhancement is instrumental in both understanding the properties of black holes
 as well as in AdS/CFT \cite{maldacena}. In the context of black holes, the presence of additional supersymmetries near the horizon  would imply the existence of additional isometries as well as the presence of
 fundamental forms associated with a more refined $G$-structure than that associated with a single Killing spinor.  This typically leads to more geometric restrictions on the horizon sections which in turn may lead  to the classification of  the  horizons which preserve sufficiently large number of supersymmetries. For some historical references on uniqueness theorems of black holes and some new results see \cite{israel}-\cite{ring}. In the context of  AdS/CFT, the enhancement of the bosonic symmetries to $\mathfrak{sl}(2, \mathbb{R})$ is the minimal required to assert that the dual field theory is conformal.

  In this paper, we shall demonstrate that M-horizons with smooth fields\footnote{The near horizon geometry of NS5-branes preserves the same number of supersymmetries as the NS5-brane but the dilaton is not constant at the lightcone.} preserve an even number of supersymmetries. The geometry of M-horizons that preserve one supersymmetry have been investigated in \cite{smhor, mhor}. Here we shall show that they admit a second supersymmetry. Our proof\footnote{We do not assume the bilinear matching condition, ie the
   identification of the stationary Killing vector field of the black hole with the vector constructed as a Killing spinor bilinear, which has  extensively been used in most of the literature
   on near horizon geometries. However see \cite{4dhor}.}  is topological in nature, and utilizes in an essential way the field
   equations of 11-dimensional supergravity and
  the properties of two horizon Dirac operators ${\cal D}^{(\pm)}$. The horizon Dirac operators are defined on the horizon sections, which are compact 9-dimensional manifolds without boundary,
 and are constructed from the supercovariant connection of 11-dimensional supergravity  after integrating out the lightcone directions.  As a result, they exhibit couplings associated to the 4-form
  field strength of 11-dimensional supergravity,  and  arise naturally from the
  investigation of the Killing spinor equations (KSEs) for near horizon geometries.
   Our proof that the number of supersymmetries of near horizon geometries is even  proceeds with establishing of Lichnerowicz type theorems for both horizon Dirac operators.
   These theorems relate the number of Killing spinors to the zero modes of the horizon Dirac operators and they are valid subject to the field equations
   and Bianchi identities of 11-dimensional supergravity. The number
   of supersymmetries, $N>0$, of a near horizon geometry is given by $N=\mathrm {dim~ Ker}\, {\cal D}^{(+)}+\mathrm {dim~ Ker}\, {\cal D}^{(-)}$.
   The proof continues by utilizing the vanishing of the index  of   Dirac operators on odd dimensional manifolds which can be used to demonstrate that $\mathrm{dim~ Ker}\, {\cal D}^{(+)}=\mathrm {dim~ Ker}\, {\cal D}^{(-)}$.  This in turn implies that the number of supersymmetries preserved by near horizon geometries is even.

   Our theorem implies that the near horizon geometries of 11-dimensional supergravity preserve at least two supersymmetries.  We shall demonstrate that a consequence   of this is that all near horizon geometries of 11-dimensional supergravity with non-trivial fluxes admit a $\mathfrak{sl}(2, \mathbb{R})$
   subgroup of isometries. The orbits of the $\mathfrak{sl}(2, \mathbb{R})$ symmetry  on the near horizon spacetime are either 2- or 3-dimensional. We find that if the orbits are only 2-dimensional, then the near horizon
    geometry is static and it is a warped product of $AdS_2$ with the near horizon section ${\cal S}$.  Static M-horizons have been investigated before in \cite{smhor}  and are related $AdS_2$ backgrounds
    in M-theory initiated in \cite{kim}.

   To prove our results, we have used details of the supersymmetry transformations, field
   equations and Bianchi identities of 11-dimensional supergravity. However, the methodology
   used can be applied to all  supergravity theories. Therefore, it is likely that all
   near horizon geometries of odd-dimensional supergravity theories preserve at least two
   supersymmetries and their symmetry algebra includes $\mathfrak{sl}(2, \mathbb{R})$.  This assertion
   is supported by a similar result demonstrated for the horizons of 5-dimensional minimal
   gauged supergravity in \cite{m5dhor}.

This paper has been organized as follows. In section 2, we describe the field content of M-horizons and express the Bianchi identities and
field equations of the theory on the horizon sections. In section 3, we integrate the KSEs along the lightcone directions and establish
the independent KSEs on the horizon sections. In section 4, we prove the Lichnerowicz type theorems for two horizon Dirac operators and
explore  the index theorem for the Dirac operator to prove that the number of supersymmetries is even. In section 5, we investigate further the Killing spinors of M-horizons. In section 6, we explore the consequences for the geometry and topology of M-horizons to admit at least
two supersymmetries and demonstrate that all M-horizons admit an $\mathfrak{sl}(2,\mathbb{R})$ symmetry subalgebra. In section 7, we give our conclusions.  In appendices A and B, we give more details
about the proof of the Lichnerowicz type theorems we use, and present an alternative proof
of one of the Lichnerowicz type  theorems using the maximum principle, respectively.

\section{Near Horizon Geometry}

\subsection{Near horizon fields }

Adapting Gaussian null coordinates\cite{isen, gnull},  the near horizon metric and 4-form field strength\footnote{Let $\omega$ be a k-form, then
$d_h\omega= d\omega-h\wedge \omega$.} of 11-dimensional supergravity can be written \cite{smhor, mhor} as \begin{eqnarray}
ds^2 &=& 2 {\bf{e}}^+ {\bf{e}}^- + \delta_{ij} {\bf{e}}^i {\bf{e}}^j=2 du (dr + r h - {1 \over 2} r^2 \Delta du)+ ds^2({\cal S})~,
\cr
F &=& {\bf{e}}^+ \wedge {\bf{e}}^- \wedge Y
+ r {\bf{e}}^+ \wedge d_h Y + X~,
\label{mhm}
\end{eqnarray}
where we have introduced the frame
\begin{eqnarray}
{\bf{e}}^+ = du~,~~~{\bf{e}}^- = dr + r h - {1 \over 2} r^2 \Delta du~,~~~
{\bf{e}}^i = e^i{}_J dy^J~;~~~g_{IJ}=\delta_{ij} e^i{}_I e^j{}_J~,
\label{nhbasis}
\end{eqnarray}
and
\begin{eqnarray}
ds^2({\cal S})=\delta_{ij} {\bf{e}}^i{\bf{e}}^j~,
\end{eqnarray}
is the metric of the horizon section ${\cal S}$ given by $r=u=0$. ${\cal S}$ is taken to be compact, connected and without boundary.
The dependence on the coordinates $r,u$ is given explicitly. In particular,  $h=h_i {\bf{e}}^i$, $\Delta$, ${\bf{e}}^i$, $Y$ and $X$ depend only on  the coordinates $y$
of ${\cal S}$.
We choose the frame indices  $i=1, 2, 3, 4, 6, 7, 8, 9, \sharp$ and we follow the conventions of \cite{system11}.
Observe that the Killing vector field $\partial_u$ is non-space-like everywhere as $\Delta\geq 0$,  and becomes null at $r=0$. There is no loss of generality
in taking $\Delta\geq 0$ as this is implied by the KSEs.

Regularity of the horizon requires that  $\Delta$,  $h$, $Y$ and $X$ are globally defined and smooth 0-, 1-, 2- and 4-forms on ${\cal S}$, respectively. This is our smoothness assumption and it is required
 to establish our result.

\subsection{Bianchi identities and field equations}

The Bianchi identities and field equations of 11-dimensional supergravity \cite{julia} can be decomposed along the lightcone directions and those of the horizon section ${\cal S}$. For the Bianchi
identity of $F$, $dF=0$, such a decomposition yields
\begin{eqnarray}
\label{clos}
dX=0 \ ,
\end{eqnarray}
ie $X$ is a closed form on ${\cal S}$.

Similarly, the field equation of the 3-form gauge potential is
\begin{eqnarray}
d \star_{11}  F -{1 \over 2} F \wedge F=0~,
\end{eqnarray}
where $\star_{11}$ is the Hodge star operation of 11-dimensional spacetime, which  can be decomposed as
\begin{eqnarray}
\label{geq1}
-\star_9 d_hY  - h \wedge \star_9 X + d \star_9 X = Y \wedge X~,
\end{eqnarray}
and
\begin{eqnarray}
\label{geq2}
-d \star_{9} Y = {1 \over 2} X \wedge X~,
\end{eqnarray}
where $\star_{9}$ is the Hodge star operation on ${\cal S}$. The  spacetime volume form  is chosen as $\epsilon_{11}= {\bf{e}}^+ \wedge {\bf{e}}^- \wedge \epsilon_{{\cal{S}}}$, where  $\epsilon_{{\cal{S}}}$
is the volume form of ${\cal S}$. Equivalently, in components, one has
\begin{eqnarray}
{\tilde{\nabla}}^i X_{i \ell_1 \ell_2 \ell_3} + 3 {\tilde{\nabla}}_{[\ell_1} Y_{\ell_2 \ell_3]}
=3 h_{[\ell_1} Y_{\ell_2 \ell_3]} + h^i X_{i \ell_1 \ell_2 \ell_3} -
{1 \over 48} \epsilon_{\ell_1 \ell_2 \ell_3}{}^{q_1 q_2 q_3 q_4 q_5 q_6}
Y_{q_1 q_2} X_{q_3 q_4 q_5 q_6} \ \
\end{eqnarray}
and
\begin{eqnarray}
\label{ydiv}
{\tilde{\nabla}}^j Y_{ji}-{1\over 1152} \epsilon_{i}{}^{q_1 q_2 q_3 q_4 q_5 q_6 q_7 q_8} X_{q_1 q_2 q_3 q_4} X_{q_5 q_6 q_7 q_8}=0~,
\end{eqnarray}
where ${\tilde{\nabla}}$ is the Levi-Civita connection of the metric $ds^2({\cal S})$ on the near horizon section ${\cal{S}}$.

The Einstein equation is
\begin{eqnarray}
R_{MN} = {1 \over 12} F_{M L_1 L_2 L_3} F_N{}^{L_1 L_2 L_3}
-{1 \over 144} g_{MN} F_{L_1 L_2 L_3 L_4}F^{L_1 L_2 L_3 L_4} \ .
\end{eqnarray}
This decomposes into a number of components. In particular along ${\cal{S}}$, one finds
\begin{eqnarray}
\label{ein1}
{\tilde {R}}_{ij} + {\tilde{\nabla}}_{(i} h_{j)} -{1 \over 2} h_i h_j &=& -{1 \over 2} Y_{i \ell} Y_j{}^\ell
+{1 \over 12} X_{i \ell_1 \ell_2 \ell_3} X_j{}^{\ell_1 \ell_2 \ell_3}
\nonumber \\
&+& \delta_{ij} \bigg( {1 \over 12} Y_{\ell_1 \ell_2} Y^{\ell_1 \ell_2}
-{1 \over 144} X_{\ell_1 \ell_2 \ell_3 \ell_4} X^{\ell_1 \ell_2 \ell_3 \ell_4} \bigg)~,
\end{eqnarray}
where  ${\tilde{R}}_{ij}$ is the Ricci tensor of ${\cal{S}}$.
The $+-$ component of the Einstein equation gives
\begin{eqnarray}
\label{einpm}
{\tilde{\nabla}}^i h_i = 2 \Delta + h^2 -{1 \over 3} Y_{\ell_1 \ell_2} Y^{\ell_1 \ell_2}
-{1 \over 72} X_{\ell_1 \ell_2 \ell_3 \ell_4} X^{\ell_1 \ell_2 \ell_3 \ell_4}~.
\end{eqnarray}
Similarly, the $++$ and $+i$ components of the Einstein equation can be expressed as
\begin{eqnarray}
\label{einpp}
{1 \over 2} {\tilde{\nabla}}^i {\tilde{\nabla}}_i \Delta -{3 \over 2} h^i {\tilde{\nabla}}_i \Delta -{1 \over 2} \Delta
{\tilde{\nabla}}^i h_i + \Delta h^2 +{1 \over 4} dh_{ij} dh^{ij}
= {1 \over 12} (d_hY)_{\ell_1 \ell_2 \ell_3} (d_h Y)^{\ell_1 \ell_2 \ell_3}~,
\nonumber \\
\end{eqnarray}
and
\begin{eqnarray}
\label{einpi}
-{1 \over 2} {\tilde{\nabla}}^j dh_{ji} + h^j (dh)_{ji} - {\tilde{\nabla}}_i \Delta + \Delta h_i
= {1 \over 12} X_i{}^{\ell_1 \ell_2 \ell_3} (d_h Y)_{\ell_1 \ell_2 \ell_3}
-{1 \over 4} (d_h Y)_i{}^{\ell_1 \ell_2} Y_{\ell_1 \ell_2} \ ,
\nonumber \\
\end{eqnarray}
respectively.
Although we have included the $++$ and the $+i$ components of the
Einstein equations for completeness, it is straightforward to show that
both ({\ref{einpp}}) and ({\ref{einpi}}) hold as a consequence of
({\ref{clos}}), the 3-form field equations ({\ref{geq1}}) and ({\ref{geq2}}) and
the components of the Einstein equation in ({\ref{ein1}}) and  ({\ref{einpm}}).
This does not make use of supersymmetry, or any assumptions on the topology of
${\cal{S}}$.
Hence, the conditions on $ds^2({\cal S})$,  $\Delta$, $h$, $Y$ and $X$ simplify to
({\ref{clos}}), ({\ref{geq1}}), ({\ref{geq2}}), ({\ref{ein1}}) and ({\ref{einpm}}).

\section{Killing spinor equations}

The KSE of 11-dimensional supergravity \cite{julia} is
\begin{eqnarray}
\nabla_M \epsilon
+\bigg(-{1 \over 288} \Gamma_M{}^{L_1 L_2 L_3 L_4} F_{L_1 L_2 L_3 L_4}
+{1 \over 36} F_{M L_1 L_2 L_3} \Gamma^{L_1 L_2 L_3} \bigg) \epsilon =0~,
\nonumber \\
\end{eqnarray}
where $\nabla$ is the spacetime Levi-Civita connection.  As for the field equations above, the KSE can be decomposed along the
light-cone and ${\cal S}$ directions. This already has been done in \cite{smhor} and \cite{mhor}. Here we shall repeat
some of the steps as our analysis is different from that in the above references.  In particular, we shall not assume that there is
bi-linear matching, ie that the Killing vector field  $\partial_u$ is identified with the vector constructed as bi-linear of the Killing spinor  of backgrounds
preserving one supersymmetry.

\subsection{Integration of KSEs along the lightcone}

To solve the KSEs along the light-cone directions, we decompose the Killing spinor as
\begin{eqnarray}
\epsilon = \epsilon_+ + \epsilon_-~,~~~\Gamma_\pm \epsilon_\pm =0~.
\label{dec}
\end{eqnarray}
Then after some computation, see \cite{smhor, mhor}, we find that
\begin{eqnarray}
\label{ksp1}
\epsilon_+ = \eta_+, \qquad \epsilon_- = \eta_- + r \Gamma_-
\Theta_+\eta_+~,
\end{eqnarray}
and
\begin{eqnarray}
\label{ksp2}
\eta_+ = \phi_+ + u \Gamma_+ \Theta_- \phi_- , \qquad \eta_- = \phi_-~,
\end{eqnarray}
where
\begin{eqnarray}
\Theta_\pm=\bigg({1 \over 4} h_i \Gamma^i +{1 \over 288} X_{\ell_1 \ell_2 \ell_3 \ell_4}
\Gamma^{\ell_1 \ell_2 \ell_3 \ell_4} \pm {1 \over 12} Y_{\ell_1 \ell_2} \Gamma^{\ell_1 \ell_2} \bigg)~,
\end{eqnarray}
and $\phi_\pm = \phi_\pm (y)$ do not depend on $r$ or $u$.

Furthermore, the $+$ and $-$
components of the KSE impose the following algebraic conditions on the Killing spinors
\begin{eqnarray}
\label{cc1}
&&\bigg( {1 \over 2} \Delta -{1 \over 8} dh_{ij} \Gamma^{ij} +{1 \over 72} d_hY_{\ell_1 \ell_2 \ell_3}
\Gamma^{\ell_1 \ell_2 \ell_3}
\nonumber \\
&+&2 \big({1 \over 4} h_i \Gamma^i -{1 \over 288} X_{\ell_1 \ell_2 \ell_3 \ell_4}
\Gamma^{\ell_1 \ell_2 \ell_3 \ell_4} +{1 \over 12} Y_{\ell_1 \ell_2} \Gamma^{\ell_1 \ell_2} \big)
\Theta_+ \bigg) \eta_+ =0~,
\end{eqnarray}
\begin{eqnarray}
\label{cc2}
&& \bigg({1 \over 4} \Delta h_i \Gamma^i - {1 \over 4} \partial_i \Delta \Gamma^i
+ \big( -{1 \over 8} dh_{ij} \Gamma^{ij} -{1 \over 24} d_hY_{\ell_1 \ell_2 \ell_3} \Gamma^{\ell_1 \ell_2
\ell_3} \big)\,
\Theta_+ \bigg) \eta_+ =0~,
\end{eqnarray}
and
\begin{eqnarray}
\label{cc3}
&& \bigg( -{1 \over 2} \Delta -{1 \over 8} dh_{ij} \Gamma^{ij} +{1 \over 24}
d_hY_{\ell_1 \ell_2 \ell_3} \Gamma^{\ell_1 \ell_2 \ell_3}
\nonumber \\
&+& 2 \big(-{1 \over 4} h_n \Gamma^n +{1 \over 288} X_{n_1 n_2 n_3 n_4}
\Gamma^{n_1 n_2 n_3 n_4} +{1 \over 12} Y_{n_1 n_2} \Gamma^{n_1 n_2} \big)
\Theta_- \bigg)
\phi_- =0~.
 \end{eqnarray}
Note that conditions ({\ref{cc1}}) and ({\ref{cc3}}) can be expanded out into terms independent
of $u$ and terms linear in $u$, yielding four $u$-independent conditions. However,
it will turn out to be most convenient to write the algebraic conditions in the form
of ({\ref{cc1}}) and ({\ref{cc2}}).

Since we have separated the light-cone directions from the rest, the remaining KSEs have manifest $Spin(9)\subset Spin(10,1)$ local gauge
invariance. The remaining KSEs can be written as

\begin{eqnarray}
\label{sp1}
{\tilde{\nabla}}_i \eta_+ + \bigg( -{1 \over 4} h_i -{1 \over 288} \Gamma_i{}^{\ell_1 \ell_2 \ell_3 \ell_4}
X_{\ell_1 \ell_2 \ell_3 \ell_4} +{1 \over 36} X_{i \ell_1 \ell_2 \ell_3} \Gamma^{\ell_1 \ell_2 \ell_3}
\nonumber \\
+{1 \over 24} \Gamma_i{}^{\ell_1 \ell_2} Y_{\ell_1 \ell_2} -{1 \over 6} Y_{ij}\Gamma^j \bigg) \eta_+ =0~,
\end{eqnarray}
\begin{eqnarray}
\label{sp2}
{\tilde{\nabla}}_i \zeta +  \bigg(-{1 \over 2} h_i +{1 \over 4} \Gamma_i{}^\ell h_\ell
-{1 \over 24} X_{i \ell_1 \ell_2 \ell_3} \Gamma^{\ell_1 \ell_2 \ell_3}
+{1 \over 8} \Gamma_i{}^{\ell_1 \ell_2} Y_{\ell_1 \ell_2} \bigg) \zeta
\nonumber \\
+ \bigg( {1 \over 4} \Delta \Gamma_i -{1 \over 16} \Gamma_i{}^{\ell_1 \ell_2} dh_{\ell_1 \ell_2}
-{3 \over 8} dh_{i \ell}\Gamma^{\ell} -{1 \over 48} d_hY_{\ell_1 \ell_2 \ell_3}
\Gamma^{\ell_1 \ell_2 \ell_3} \Gamma_i \bigg) \eta_+ =0~,
\nonumber \\
\end{eqnarray}
and
\begin{eqnarray}
\label{sp3}
{\tilde{\nabla}}_i \phi_- + \bigg( {1 \over 4} h_i -{1 \over 288} \Gamma_i{}^{\ell_1 \ell_2 \ell_3 \ell_4}
X_{\ell_1 \ell_2 \ell_3 \ell_4} +{1 \over 36} X_{i \ell_1 \ell_2 \ell_3} \Gamma^{\ell_1 \ell_2 \ell_3}
\nonumber \\
-{1 \over 24} \Gamma_i{}^{\ell_1 \ell_2} Y_{\ell_1 \ell_2} +{1 \over 6} Y_{ij}\Gamma^j \bigg) \phi_-=0~.
\end{eqnarray}
where
\begin{eqnarray}
\zeta \equiv \Theta_+ \eta_+~.
\end{eqnarray}

Again, we remark that equations ({\ref{sp1}}) and ({\ref{sp2}}) contain both terms independent of
$u$, and linear in $u$, but it is most convenient to write these equations in the way stated above.

\subsection{Independent KSEs on ${\cal S}$}
\label{ppp}

It is important to what follows to establish the independent KSEs on ${\cal S}$. In investigating supersymmetric backgrounds,
it is customary to first solve all KSEs and then impose the field equations which are not implied as integrability conditions.  Here,
we shall adopt a different strategy.  We shall use all the field equations and Bianchi identities of the theory to find the independent
KSEs that one has to impose such that a near horizon geometry is supersymmetric.

First consider equations ({\ref{cc1}}), ({\ref{cc2}}), ({\ref{sp1}}) and ({\ref{sp2}}).
Equations of this form have already been investigated in \cite{mhor},
in the special case for which $\phi_-=0$. However, the form of these equations
remains unchanged if one relaxes the condition that $\phi_-$ vanishes.  Hence
using exactly the same reasoning set out in \cite{mhor}, it follows that
({\ref{cc1}}), ({\ref{cc2}}) and ({\ref{sp2}}) are implied by ({\ref{sp1}}) and the
bosonic field equations and Bianchi identities.

Next we consider ({\ref{sp1}}) in more detail. It will be particularly useful to establish the following
result: if $\phi_-$ is a ($u, r$-independent)  spinor satisfying ({\ref{sp3}}),
then
\begin{eqnarray}
\label{etdef}
\phi_+' \equiv \Gamma_+ \Theta_- \phi_-
\end{eqnarray}
satisfies ({\ref{sp1}}) with $\eta_+$ replaced with $\phi_+'$.
To see this, first evaluate the LHS of ({\ref{sp1}}) acting on $\eta_+'$, and use ({\ref{sp3}}) to eliminate the terms involving ${\tilde{\nabla}}_i \phi_-$, and then remove the $\Gamma_+$ term by
left-multiplication with $\Gamma_-$.
Then compare the resulting algebraic condition on $\phi_-$ with the following expression
\begin{eqnarray}
\label{identaux1}
{1 \over 2} \Gamma^j ({\tilde{\nabla}}_j {\tilde{\nabla}}_i - {\tilde{\nabla}}_i {\tilde{\nabla}}_j) \phi_- = {1 \over 4} {\tilde{R}}_{ij}
\Gamma^j \phi_-~,
\end{eqnarray}
where the LHS of the above is evaluated using ({\ref{sp3}}), and the entire expression
is then simplified using the bosonic field equations and Bianchi identities.
After a rather involved computation, one finds that the resulting algebraic
condition on $\phi_-$ obtained from ({\ref{identaux1}}) is identical to the algebraic
condition on $\phi_-$ obtained by substituting $\phi_+'$ into ({\ref{sp1}}) as described above.
We remark that the condition ({\ref{cc3}}) was not used at any stage of the computation.
Note that this result implies that the part of ({\ref{sp1}}) which is linear in $u$ is satisfied
automatically as a consequence of ({\ref{sp3}}) and the field equations and Bianchi identities.

Next consider ({\ref{cc3}}). It will again be useful to return to ({\ref{identaux1}}),
with the LHS evaluated using ({\ref{sp3}}). On contracting the resulting
condition with $\Gamma^i$ and making extensive use of the field equations and Bianchi identities to simplify the expression, one obtains a condition equivalent to ({\ref{cc3}}).

To summarize, we have demonstrated that on making use of the field equations and Bianchi identities the
independent KSEs are
\begin{eqnarray}
\label{ind1}
\nabla_i^{(+)}\phi_+\equiv {\tilde{\nabla}}_i \phi_+ + \Psi^{(+)}_i \phi_+ =0~,
\end{eqnarray}
and
\begin{eqnarray}
\label{ind2}
\nabla_i^{(-)}\phi_-\equiv {\tilde{\nabla}}_i \phi_- + \Psi^{(-)}_i \phi_- =0~,
\end{eqnarray}
where
\begin{eqnarray}
\Psi^{(\pm)}_i &=& \mp{1 \over 4} h_i -{1 \over 288} \Gamma_i{}^{\ell_1 \ell_2 \ell_3 \ell_4}
X_{\ell_1 \ell_2 \ell_3 \ell_4} +{1 \over 36} X_{i \ell_1 \ell_2 \ell_3} \Gamma^{\ell_1 \ell_2 \ell_3}
\nonumber \\
&\pm&{1 \over 24} \Gamma_i{}^{\ell_1 \ell_2} Y_{\ell_1 \ell_2} \mp{1 \over 6} Y_{ij} \Gamma^j~.
\end{eqnarray}
Moreover on imposing the field equations and Bianchi identities,  if $\phi_-$ satisfies ({\ref{ind2}}), then $\phi_+'$ given by
({\ref{etdef}}) satisfies ({\ref{ind1}}).
This analysis has been entirely local, and has not made use of the compactness of ${\cal{S}}$.

\section{Horizon Dirac Equations and a Lichnerowicz Theorem}

\subsection{Horizon Dirac Equations}

Given the gravitino KSE in a supergravity theory which is  a parallel transport equation for the supercovariant connection, ${\cal D}_A \epsilon =0$, one can construct a ``supergravity Dirac equation'' as $\Gamma^A {\cal D}_A \epsilon=0$. This can be adapted to the near horizon geometries. In particular, for each of the ``horizon gravitino KSEs'' on ${\cal S}$
\begin{eqnarray}
\label{redkse1}
\nabla_i^{(\pm)}\phi_\pm \equiv {\tilde{\nabla}}_i \phi_\pm + \Psi^{(\pm)}_i \phi_\pm =0~,
\end{eqnarray}
 given in ({\ref{ind1}}) and  ({\ref{ind2}}), respectively,
one can associate a``horizon Dirac equation''  as
\begin{eqnarray}
\label{dirac1}
{\cal{D}}^{(\pm)}\phi_\pm=\Gamma^i {\tilde{\nabla}}_i \phi_\pm + \Psi^{(\pm)} \phi_\pm =0 \ ,
\end{eqnarray}
where
\begin{eqnarray}
\Psi^{(\pm)} = \Gamma^i \Psi^{(\pm)}_i = \mp{1 \over 4} h_\ell \Gamma^\ell +{1 \over 96} X_{\ell_1 \ell_2 \ell_3 \ell_4}
\Gamma^{\ell_1 \ell_2 \ell_3 \ell_4} \pm{1 \over 8} Y_{\ell_1 \ell_2} \Gamma^{\ell_1 \ell_2}~.
\end{eqnarray}
These Dirac equations,  in addition to the Levi-Civita connection, also depend on the fluxes of the supergravity theory restricted on the horizon section ${\cal S}$.

\subsection { A  Lichnerowicz theorem}

The horizon  Dirac equations (\ref{dirac1})  can be used to give a new characterization of the Killing spinors.
Clearly, the gravitino KSEs are more restrictive.  Any solution  of the gravitino KSEs ({\ref{ind1}}), ({\ref{ind2}}) is also a solution
of a corresponding Dirac equation. In what follows, we shall explore the converse. In particular, we shall show that
the zero modes of the horizon Dirac equation are parallel with respect to the horizon supercovariant derivatives (\ref{redkse1}). Instrumental in the proof
are the field equations and Bianchi identities of 11-dimensional supergravity as reduced on the horizon section ${\cal S}$.

Before proceeding with the analysis of the supergravity case, it is useful to recall the Lichnerowicz theorem. On any spin compact manifold $N$, one can show the equality
\begin{eqnarray}
\int_N \langle \Gamma^i \nabla_i \epsilon, \Gamma^j \nabla_j \epsilon \rangle=  \int_N \langle  \nabla_i \epsilon ,  \nabla^i \epsilon \rangle+\int_N {R\over 4} \langle \epsilon , \epsilon \rangle~,
\end{eqnarray}
where $\nabla$ is the Levi-Civita connection, $\langle \cdot, \cdot\rangle$ is the Dirac inner product and $R$ is the Ricci scalar. Clearly if $R>0$, the Dirac operator has no zero modes. Moreover, if
$R=0$, then the zero modes of the Dirac operator are parallel with respect to the Levi-Civita connection.

This theorem can be generalized for M-horizons with the standard Dirac operator replaced with the horizon Dirac operators in (\ref{dirac1}) and the Levi-Civita covariant derivative
replaced with the horizon supercovariant derivatives  (\ref{redkse1}). A version of this theorem has already been proven in \cite{mhor} but here
we shall consider both horizon Dirac operators ${\cal D}^{(\pm)}$. For this, let $\phi$ be a Majorana $Spin(9)$ spinor and consider
\begin{eqnarray}
\label{lich1}
{\cal{I}}^{(\pm)} = \int_{\cal{S}} \langle {\tilde{\nabla}}_i \phi_\pm + \Psi^{(\pm)}_i \phi_\pm, {\tilde{\nabla}}^i \phi_\pm + \Psi^{{(\pm)}i} \phi_\pm\rangle
- \int_{\cal{S}} \parallel\Gamma^i {\tilde{\nabla}}_i \phi_\pm + \Psi^{(\pm)} \phi_\pm\parallel^2~,
\end{eqnarray}
where $ \langle\cdot, \cdot \rangle$ is the Dirac inner product\footnote{In fact, it can be identified
 as the restriction of the $Spin(10,1)$ invariant inner product on the Majorana representation
 as restricted on $Spin(9)$ spinor representations under the decomposition $\epsilon=\epsilon_++\epsilon_-$ in (\ref{dec}).} of $Spin(9)$ which can be identified
with the standard Hermitian inner product on $\Lambda^*(\mathbb{C}^4)$ restricted on the real subspace of Majorana spinors and $\parallel \cdot \parallel$ is the associated norm. Therefore, $ \langle\cdot, \cdot \rangle$ is a real and positive definite. The   $Spin(9)$ gamma matrices are Hermitian with respect to
$ \langle\cdot, \cdot \rangle$.

Clearly, if the integrals ${\cal{I}}^{(\pm)}$ vanish, all zero modes of the horizon Dirac operators
${\cal D}^\pm$ are parallel with respect to the horizon supercovariant derivatives $\nabla^\pm$ and so Killing. To show that ${\cal{I}}^{(\pm)}$ vanish, assume that ${\cal{S}}$ is compact and without boundary, and that  $\phi$ is globally well-defined
and smooth on ${\cal{S}}$.
Then, on integrating by parts, one can rewrite
\begin{eqnarray}
\label{lichaux}
{\cal{I}}^{(\pm)} &=& \int_{\cal{S}} \langle \phi_\pm, (\Psi^{(\pm)i \dagger} - \Psi^{(\pm)i}-(\Psi^{(\pm)\dagger}-\Psi^{(\pm)})\Gamma^i) {\tilde{\nabla}}_i \phi_\pm
+ (\Psi^{(\pm)}{}_i^\dagger \Psi^{(\pm)i} - \Psi^{(\pm) \dagger} \Psi^{(\pm)}) \phi_\pm
\nonumber \\
&+& \Gamma^{ij} {\tilde{\nabla}}_i {\tilde{\nabla}}_j \phi_\pm
+(\Gamma^i {\tilde{\nabla}}_i \Psi^{(\pm)} - ({\tilde{\nabla}}^i \Psi^{(\pm)}_i)) \phi_\pm + (\Gamma^i \Psi^{(\pm)}- \Psi^{(\pm)} \Gamma^i) {\tilde{\nabla}}_i \phi_\pm \rangle~.
\end{eqnarray}
Next, evaluating the RHS of the above equation using the Bianchi identity of $X$ (\ref{clos}), the field equation of the 4-form field strength ({\ref{geq1}}) and ({\ref{geq2}}),
the Einstein equations along ${\cal{S}}$ ({\ref{ein1}}),
one finds that
\begin{eqnarray}
\label{lich2}
{\cal{I}}^{(\pm)} &=&  \int_{\cal{S}} \langle \phi_\pm , \big( \mp{1 \over 2} h_\ell \Gamma^\ell
-{1 \over 144} X_{\ell_1 \ell_2 \ell_3 \ell_4} \Gamma^{\ell_1 \ell_2 \ell_3 \ell_4}
\mp {1 \over 6} Y_{\ell_1 \ell_2} \Gamma^{\ell_1 \ell_2} \big)
(\Gamma^i {\tilde{\nabla}}_i \phi_\pm + \Psi^{(\pm)} \phi_\pm) \rangle
\nonumber \\
&+& \int_{\cal{S}} {1 \over 2} {\tilde{\nabla}}^i h_i (1 \pm 1) \langle \phi_\pm
, \phi_\pm \rangle~.
\end{eqnarray}
Further details of this computation are given in Appendix A.

On comparing ({\ref{lich2}}) with ({\ref{lich1}}), one immediately finds
that if $\phi_-$ is a solution of the ${\cal D}^{(-)}$ horizon Dirac equation, then $\phi_-$ is a solution of the $\nabla^{(-)}$ horizon gravitino KSE ({\ref{ind2}}).

Also, if $\phi_+$ is a solution of the ${\cal D}^{(+)}$ horizon Dirac equation, and
$\langle \phi_+ , \phi_+ \rangle =\mathrm{const}$ then   $\phi_+$ is a solution
of the $\nabla^{(+)}$ horizon graviton KSE ({\ref{ind1}}). Furthermore, it is straightforward to
prove that if $\phi_+$ satisfies the ${\cal D}^{(+)}$ horizon Dirac equation, then
$\langle \phi_+ , \phi_+ \rangle = \mathrm{const}$ follows from an application of the
maximum principle. In particular, if one assumes that ${\cal{D}}^{(+)} \phi_+ =0$, then
one obtains the condition
\begin{eqnarray}
\label{max2}
{\tilde{\nabla}}^i {\tilde{\nabla}}_i \parallel \phi_+\parallel^2 -h^i {\tilde{\nabla}}_i \parallel \phi_+\parallel^2
= 2 \langle {\tilde{\nabla}}^{(+)}{}^i \phi_+ , {\tilde{\nabla}}^{(+)}_i \phi_+ \rangle~.
\end{eqnarray}
Details of the derivation of ({\ref{max2}}) are given in Appendix B.
Upon using the maximum principle, this condition implies that $\parallel\phi_+\parallel = \mathrm{const}$,
and ${\tilde{\nabla}}^{(+)}_i \phi_+ =0$. Note that this provides an alternative proof of the Lichnerowicz type of theorem
for the ${\cal D}^{(+)}$ operator.

To conclude,  the results of this section can be summarized as
\begin{eqnarray}
\nabla^{(\pm)}_i \phi_\pm=0\Longleftrightarrow {\cal D}^{(\pm)} \phi_\pm=0~.
\end{eqnarray}
Hence, the Killing spinors of the horizon section ${\cal S}$ can be identified with the zero modes of horizon Dirac operators. In turn, the Killing spinors of the near horizon spacetime can be expressed
in terms of the zero models of the horizon Dirac operators ${\cal D}^{(\pm)}$.

\subsection{Index theorem and  supersymmetries of M-horizons}

To proceed note that we have decomposed the spin bundle $S$ of 11-dimensional supergravity as $S=S_+\oplus S_-$ on ${\cal S}$ using the
 projections $\Gamma_\pm$ as in (\ref{dec}). Next observe that ${\cal D}^{(+)}: ~\Gamma(S_+)\rightarrow \Gamma(S_+)$ and its adjoint  $({\cal D}^{(+)})^\dagger: ~\Gamma(S_+)\rightarrow \Gamma(S_+)$, where
 $\Gamma(S_+)$ are the smooth sections of $S_+$.  The operator  ${\cal D}^{(+)}$ has the same principal symbol as the Dirac operator. Moreover , ${\cal D}^{(+)}$ is defined on  ${\cal{S}}$ which is an odd-dimensional manifold.
It follows from Proposition 1 of \cite{atiyah1} that the index of ${\cal D}^{(+)}$ vanishes. As a result, we have that
\begin{eqnarray}
{\rm dim}\, {\rm ker} {\cal D}^{(+)}  = {\rm dim}\, {\rm ker} ({\cal D}^{(+)})^\dagger ~.
\label{index}
\end{eqnarray}
Observe that
\begin{eqnarray}
({\cal D}^{(+)})^\dagger= -\Gamma^i{\tilde{\nabla}}_i-{1 \over 4} h_\ell \Gamma^\ell +{1 \over 96} X_{\ell_1 \ell_2 \ell_3 \ell_4}
\Gamma^{\ell_1 \ell_2 \ell_3 \ell_4} -{1 \over 8} Y_{\ell_1 \ell_2} \Gamma^{\ell_1 \ell_2}~.
\end{eqnarray}
Next  define  ${\phi}'_+ = \Gamma_+ \phi_-$ and observe that
\begin{eqnarray}
({\cal D}^{(+)})^\dagger{\phi}'_+=\Gamma_+{\cal D}^{(-)}\phi_-~.
\end{eqnarray}
So, we conclude that $ {\rm dim}\, {\rm ker} ({\cal D}^{(+)})^\dagger={\rm dim}\, {\rm ker} {\cal D}^{(-)}$. This together with the
result from the index theorem (\ref{index})  gives
\begin{eqnarray}
{\rm dim} \, {\rm ker} {\cal{D}}^{(+)} = {\rm dim}\, {\rm ker} {\cal{D}}^{(-)}~.
\end{eqnarray}

The number of supersymmetries of a near horizon geometry is the number of parallel spinors of $\nabla^{(\pm)}$ and so from the Lichnerowicz theorems and the above
formula, one has
\begin{eqnarray}
N={\rm dim} \, {\rm ker} {\cal{D}}^{(+)} + {\rm dim} {\rm ker} {\cal{D}}^{(-)}=2\, {\rm dim}\, {\rm ker} {\cal{D}}^{(-)}.
\end{eqnarray}
This proves that the number of supersymmetries preserved by near M-horizon geometries is even.

\section{Construction of $\phi_+$ from $\phi_-$ Killing spinors}

In section (\ref{ppp}), we have demonstrated that if $\phi_-$  is ${\tilde{\nabla}}^{(-)}$-parallel,  then
 \begin{eqnarray}
\label{extragen}
\phi_+ = \Gamma_+ \Theta_- \phi_-
\end{eqnarray}
satisfies ${\tilde{\nabla}}^{(+)} \phi_+=0$. Clearly this can be used to construct the $\phi_+$ solutions to the KSEs from the $\phi_-$ solutions.

Since $\phi_+$ given in (\ref{extragen}) is ${\tilde{\nabla}}^{(+)}$-parallel  either is everywhere nonzero or vanishes identically. Consider the latter case  that $\phi_+$ in ({\ref{extragen}})
 vanishes for $\phi_-\not=0$. In this case, one must have
\begin{eqnarray}
\label{cond2}
\Theta_- \phi_- =0 \ .
\end{eqnarray}
To proceed, note that ({\ref{cc3}}) together with ({\ref{cond2}}) imply that
\begin{eqnarray}
\langle \phi_- ,  \bigg( -{1 \over 2} \Delta -{1 \over 8} dh_{ij} \Gamma^{ij} +{1 \over 24}
d_hY_{\ell_1 \ell_2 \ell_3} \Gamma^{\ell_1 \ell_2 \ell_3} \bigg) \phi_- \rangle =0~.
\end{eqnarray}
This condition implies
\begin{eqnarray}
\Delta \langle \phi_- , \phi_- \rangle =0
\end{eqnarray}
and hence
\begin{eqnarray}
\Delta =0 \ ,
\end{eqnarray}
as $\phi_-$ is no-where vanishing.
Next, using ({\ref{ind2}}), one finds
\begin{eqnarray}
{\tilde{\nabla}}_i \langle \phi_-, \phi_- \rangle = -{1 \over 2} h_i  \langle \phi_-, \phi_- \rangle
+ \langle \phi_- , \bigg({1 \over 144} \Gamma_i{}^{\ell_1 \ell_2 \ell_3 \ell_4}
X_{\ell_1 \ell_2 \ell_3 \ell_4}-{1 \over 3} Y_{ij} \Gamma^j \bigg) \phi_- \rangle~.
\end{eqnarray}
This expression can be further simplified, using $\Theta_- \phi_-=0$, to eliminate the
$Y$ and $X$ terms and to give
\begin{eqnarray}
\label{nrm1a}
{\tilde{\nabla}}_i \langle \phi_-, \phi_- \rangle = - h_i  \langle \phi_-, \phi_- \rangle~.
\end{eqnarray}
As $\phi_-$ is no-where zero, this implies that
\begin{eqnarray}
dh=0~,
\end{eqnarray}
and ({\ref{einpp}}) then implies that
\begin{eqnarray}
d_h Y=0~,
\end{eqnarray}
as well.
Returning to ({\ref{nrm1a}}), on taking the divergence, and using ({\ref{einpm}}) to
eliminate the ${\tilde{\nabla}}^i h_i$ term, one obtains
\begin{eqnarray}
{\tilde{\nabla}}^i {\tilde{\nabla}}_i  \langle \phi_-, \phi_- \rangle = \bigg({1 \over 3} Y_{\ell_1 \ell_2}
Y^{\ell_1 \ell_2} +{1 \over 72}X_{\ell_1 \ell_2 \ell_3 \ell_4}
X^{\ell_1 \ell_2 \ell_3 \ell_4} \bigg)  \langle \phi_-, \phi_- \rangle~.
\end{eqnarray}
On integrating both sides of this expression over ${\cal{S}}$, the contribution
from the LHS vanishes, so
\begin{eqnarray}
\int_{{\cal{S}}} \bigg({1 \over 3} Y_{\ell_1 \ell_2}
Y^{\ell_1 \ell_2} +{1 \over 72}X_{\ell_1 \ell_2 \ell_3 \ell_4}
X^{\ell_1 \ell_2 \ell_3 \ell_4} \bigg)  \langle \phi_-, \phi_- \rangle =0~.
\end{eqnarray}
Again, as $\phi_-$ is no-where vanishing, this implies that
\begin{eqnarray}
Y=0, X=0~.
\end{eqnarray}
Next, ({\ref{einpm}}) implies that
\begin{eqnarray}
{\tilde{\nabla}}^i h_i = h^2
\end{eqnarray}
and again integrating this expression over ${\cal{S}}$, the contribution from
the LHS vanishes, leading to
\begin{eqnarray}
h=0~.
\end{eqnarray}

Hence,  if $\phi_- \neq 0$,  then $\Theta_- \phi_- =0$ implies that $\Delta=0, h=0, Y=0, X=0$. In such a case, the near horizon geometry is locally isometric to $\mathbb{R}^{1,1} \times {\cal{S}}$,
where ${\cal{S}}$ is a compact 9-dimensional Ricci flat manifold. Such manifolds are classified and ${\cal S}$ is locally a product
$S^1\times X^8$, where in turn  $X^8$ is a product of holonomy $Spin(7)$, $Sp(2)$, $G_2$,  $SU(k)$, $k\leq4$,  and $\{1\}$ manifolds.

Therefore there are two possibilities. One possibility is that the conditions $\Delta=0, h=0, Y=0, X=0$ do not hold in which case the $\phi_+$ and $\phi_-$ Killing spinors are related by ({\ref{extragen}}).  This is
a consequence of the index theorem which requires the number of zero modes of ${\cal D}^{(-)}$ to be equal to those of ${\cal D}^{(+)}$.
The other possibility is  whenever $\Delta=0, h=0, Y=0, X=0$. In such a case for every $\phi_-$ spinor satisfying
${\tilde{\nabla}}^{(-)} \phi_-=0$, there is a  spinor $\phi_+$ satisfying ${\tilde{\nabla}}^{(+)}\phi_+ =0$
given by $\phi_+ = \Gamma_+ \phi_-$.  In either case, the number of Killing spinors is even.

\section{The dynamical $\mathfrak{sl}(2,\mathbb{R})$ symmetry of M-horizons}

\subsection{Killing vectors}

A priori near horizon geometries admit two Killing vector field generated by $\partial_u$ and $u\partial_u-r\partial_r$.  However all known examples
exhibit a larger symmetry algebra which always includes an $\mathfrak{sl}(2,\mathbb{R})$ subalgebra.  Here we shall prove that this is a generic
property of M-horizons with non-trivial fluxes and a direct consequence of  supersymmetry. However, it should be stressed that the $\mathfrak{sl}(2,\mathbb{R})$ symmetry is dynamical
because it emerges after using the field equations of the theory.

We have shown (\ref{dec}) that the most general Killing spinor takes the form
\begin{eqnarray}
\epsilon=\phi_++ u \Gamma_+\Theta_-\phi_-+\phi_-+r \Gamma_-\Theta_+\phi_++ r u \,\Gamma_-\Theta_+\Gamma_+\Theta_-\phi_-~.
\end{eqnarray}
The two Killing spinors of the near horizon geometries can be constructed from the
pairs $(\phi_-,0)$ and $(\phi_-, \phi_+)$ with $\phi_+=\Gamma_+ \Theta_- \phi_-$. Implementing these, we find that
\begin{eqnarray}
\epsilon_1=\phi_-+u \phi_++ru\Gamma_-\Theta_+\phi_+ ~,~~~\epsilon_2=\phi_++ r \Gamma_- \Theta_+\phi_+~.
\label{tks}
\end{eqnarray}

It can be easily shown that for any two Killing spinors $\zeta_1$ and $\zeta_2$, the 1-form bilinear\footnote{The inner product
we use to define $K$ is the Dirac inner product of $Spin(10,1)$ restricted on its Majorana representation. }
\begin{eqnarray}
K= \langle(\Gamma_+-\Gamma_-)\zeta_1, \Gamma_A \zeta_2\rangle \, e^A~,
\label{bil}
\end{eqnarray}
is associated with a Killing vector which
also preserves that 4-form field strength of 11-dimensional supergravity. In particular for the two Killing spinors
(\ref{tks}), one can construct three 1-form bi-linears.  A   substitution of (\ref{tks}) into (\ref{bil}) reveals
\begin{eqnarray}
 K_1=\langle(\Gamma_+-\Gamma_-)\epsilon_1, \Gamma_A \epsilon_2\rangle \, e^A&=& (2r \langle\Gamma_+\phi_-, \Theta_+\phi_+\rangle+  r^2 u \Delta \parallel \phi_+\parallel^2) \,{\bf{e}}^+
 \nonumber \\
 &-&2u \parallel\phi_+\parallel^2\, {\bf{e}}^-+ V_i {\bf{e}}^i~,
 \cr
 K_2=\langle(\Gamma_+-\Gamma_-)\epsilon_2, \Gamma_A \epsilon_2\rangle \, e^A&=& r^2 \Delta\parallel\phi_+\parallel^2 \,{\bf{e}}^+-2 \parallel\phi_+\parallel^2 {\bf{e}}^-~,
 \cr
 K_3=\langle(\Gamma_+-\Gamma_-)\epsilon_1, \Gamma_A \epsilon_1\rangle \, e^A&=&(2\parallel\phi_-\parallel^2+4r u \langle\Gamma_+\phi_-, \Theta_+\phi_+\rangle+ r^2 u^2 \Delta \parallel\phi_+\parallel^2) {\bf{e}}^+
 \nonumber \\
 &-&2u^2 \parallel\phi_+\parallel^2 {\bf{e}}^-+2u V_i {\bf{e}}^i~,
 \label{b1forms}
 \end{eqnarray}
where we have set
\begin{eqnarray}
\label{extraiso}
V_i =  \langle \Gamma_+ \phi_- , \Gamma_i \phi_+ \rangle~.
\end{eqnarray}
Moreover, we have used the identities
\begin{eqnarray}
- \Delta\, \parallel\phi_+\parallel^2 +4  \parallel\Theta_+ \phi_+\parallel^2 =0
\end{eqnarray}
which follows from ({\ref{cc1}}), and
\begin{eqnarray}
\label{auxcond1}
 \langle \phi_+ , \Gamma_i \Theta_+ \phi_+ \rangle  =0~,
\end{eqnarray}
which follows from $\langle \phi_+, \phi_+ \rangle = \mathrm {const}$ proved in  \cite{mhor} as a consequence of
({\ref{ind1}}) and  the compactness of ${\cal{S}}$.
By construction $K_1, K_2$ and $K_3$ are associated with vector fields which leave both the near horizon metric and
the 4-form flux (\ref{mhm}) invariant.

\subsection{The geometry of ${\cal S}$}

\subsubsection{$V\not=0$}

The symmetries generated by $K_1, K_2$ and $K_3$ restrict the geometry of ${\cal S}$. To find the restrictions on ${\cal S}$,
one decomposes the Killing condition of ${\cal L}_{K_a} g=0$ and ${\cal L}_{K_a} F=0$, $a=1,2,3$ conditions along the lightcone
and transverse directions.  After a somewhat long but straightforward computation, one finds that
\begin{eqnarray}
\tilde\nabla_{(i} V_{j)}=0~,~~~\tilde {\cal L}_Vh=0 ~,~~~\tilde {\cal L}_V\Delta=0~,~~~\tilde {\cal L}_V Y=0~,~~~\tilde {\cal L}_V X=0~.
\end{eqnarray}
Therefore, ${\cal S}$ admits an isometry generated by $V$ which leaves $h, \Delta, Y$ and $X$ invariant.

In addition, one also finds some useful identities which follow from the field equations and KSEs we have stated already. These are
\begin{eqnarray}
&&-2 \parallel\phi_+\parallel^2-h_i V^i+2 \langle\Gamma_+\phi_-, \Theta_+\phi_+\rangle=0~,~~~i_V (dh)+2 d \langle\Gamma_+\phi_-, \Theta_+\phi_+\rangle=0~,
\cr
&& 2 \langle\Gamma_+\phi_-, \Theta_+\phi_+\rangle-\Delta \parallel\phi_-\parallel^2=0~,~~~
V+ \parallel\phi_-\parallel^2 h+d \parallel\phi_-\parallel^2=0~.
\label{concon}
\end{eqnarray}
Notice that the last equality in (\ref{concon}) expresses $V$ in terms of $h$.  This is significant as it generalizes a relation
derived in the context of heterotic horizons in \cite{hethora}, \cite{hethorb} which equates $h$ with  $V$. Furthermore observe that one can show that
\begin{eqnarray}
{\cal L}_V\parallel\phi_-\parallel^2=0~.
\label{inphim}
\end{eqnarray}

The geometry of ${\cal S}$ is further restricted. The existence of a no-where vanishing spinor $\phi_-$ reduces the structure group\footnote{The isotropy group of non-trivial orbits
of $Spin(9)$ in the 16-dimensional Majorana representation is $Spin(7)$. Note that $Spin(9)/Spin(7)=S^{15}$.}
of ${\cal S}$ to $Spin(7)$. The existence of a second Killing spinor $\phi_+$ reduces the structure group further. There are various possibilities that can arise
depending in which subspace $\phi_+$ lies giving isotropy groups $Spin(7)$, $SU(4)$, $G_2$ and $SU(3)$.
There are geometric restrictions on these structures which will be explored elsewhere.

\subsubsection{$V=0$}

A special case arises whenever $V=0$. In this case, the group action generated by $K_1, K_2$ and $K_3$ has only 2-dimensional orbits. A direct substitution of this condition in (\ref{concon}) reveals that
\begin{eqnarray}
\Delta \parallel\phi_-\parallel^2=2 \parallel\phi_+\parallel^2~,~~~h=\Delta^{-1} d\Delta~.
\end{eqnarray}
Since $dh=0$ and $h$ exact such horizons are static. After a coordinate transformation $r\rightarrow \Delta r$, the near horizon geometry becomes a warped product of $AdS_2$ with ${\cal S}$, $AdS_2\times_w {\cal S}$. There are further consequences of this. For example  the ++ Einstein equation ({\ref{einpp}}) implies that
$d_h Y=0$. Static M-horizons have been extensively investigated in \cite{smhor} and they are related to
M-theory $AdS_2$ backgrounds which have been initially  explored in \cite{kim}.

\subsection{$\mathfrak{sl}(2,\mathbb{R})$ symmetry of M-horizons}

To show that all M-horizons with non-trivial fluxes admit and $\mathfrak{sl}(2,\mathbb{R})$ symmetry, we use the various identities derived in the
previous section to write the vector fields associated to the 1-forms $K_1, K_2$ and $K_3$ (\ref{b1forms}) as
\begin{eqnarray}
K_1&=&-2u \parallel\phi_+\parallel^2 \partial_u+ 2r \parallel\phi_+\parallel^2 \partial_r+ V^i \tilde \partial_i~,
\cr
K_2&=&-2 \parallel\phi_+\parallel^2 \partial_u~,
\cr
K_3&=&-2u^2 \parallel\phi_+\parallel^2 \partial_u +(2 \parallel\phi_-\parallel^2+ 4ru \parallel\phi_+\parallel^2)\partial_r+ 2u V^i \tilde \partial_i~,
\end{eqnarray}
where we have used the same symbol for the 1-forms and the associated vector fields. A direct computation then reveals using (\ref{inphim}) that
\begin{eqnarray}
[K_1,K_2]=2 \parallel\phi_+\parallel^2 K_2~,~~~[K_2, K_3]=-4 \parallel\phi_+\parallel^2 K_1  ~,~~~[K_3,K_1]=2 \parallel\phi_+\parallel^2 K_3~. \ \ \ \
\end{eqnarray}
Therefore all M-horizons with non-trivial fluxes admit an $\mathfrak{sl}(2,\mathbb{R})$ symmetry subalgebra.

Away from the fixed points of $V$, the orbits of the $\mathfrak{sl}(2,\mathbb{R})$ action
are 3-dimensional. Therefore if $V\not=0$, the $\mathfrak{sl}(2,\mathbb{R})$ action action must have some 3-dimensional orbits. If $V=0$, we have shown that $\mathfrak{sl}(2,\mathbb{R})$ has only 2-dimensional
orbits isometric to $AdS_2$.

Note also that if the fluxes are trivial, the spacetime is isometric to $\mathbb{R}^2\times {\cal S}$ and ${\cal S}$ admits at least one isometry. The symmetry group in this
case has a $\mathfrak{so}(1,1)\oplus \mathfrak{u}(1)$ subalgebra.

\section{Concluding remarks}

We have demonstrated that all M-horizons preserve an even number of supersymmetries and as a consequence of this those with non-trivial fluxes admit an $\mathfrak{sl}(2, \mathbb{R})$ subalgebra of symmetries.
To establish these results, we have shown that the KSEs of the near horizon geometries are implied
from two parallel transport equations on the horizon sections which depend on the 4-form fluxes.   Then the associated Dirac equations were considered and two Lichnerowicz type  theorems were proven which related the parallel spinors on the horizon sections with the zero modes of the associated Dirac
operators.  Then  the vanishing of the Dirac index on the 9-dimensional horizon section
led to the conclusion that M-horizons preserve an  even number of supersymmetries.  The invariance of M-horizons under a $\mathfrak{sl}(2, \mathbb{R})$ subalgebra then followed as a
consequence of the supersymmetry enhancement.

Instrumental in the proof of the above results were the field equations and Bianchi identities
of 11-dimensional supergravity. Both the supersymmetry enhancement  from one to at least two
supersymmetries as well as the presence of a $\mathfrak{sl}(2, \mathbb{R})$ invariance subalgebra of M-horizons
are dynamical, and cannot be proven without the use of field equations.

Although our calculation is based on the details of 11-dimensional supergravity, like its field content
and field equations, our methodology is general and applies to all supergravities. Therefore, it is likely
that our results generalize to all odd-dimensional supergravities  leading to the conclusion that all
odd-dimensional near horizon geometries preserve at least two supersymmetries and admit a $\mathfrak{sl}(2, \mathbb{R})$ invariance subalgebra. This assertion is further strengthen by the results
in \cite{m5dhor} where similar results were established for the horizons of 5-dimensional minimal
gauged supergravity.  Our methodology can also be adapted to investigate the symmetries of
brane horizons and $AdS$ backgrounds of odd-dimensional supergravity theories.

Our results can also be applied to even-dimensional supergravity theories.  However, there are some differences.  Assuming that the required Lichnerowicz type  theorems can be established relating
the number of Killing spinors to zero modes of Dirac operators, one does not expect that the index
of the Dirac operator vanishes on the even-dimensional horizon sections.  However, the investigation
of heterotic horizons and those of 6-dimensional supergravity in \cite{hethora}, \cite{hethorb} and \cite{6dhor}
both find that those with non-trivial fluxes  preserve an even number of supersymmetries and have
an $\mathfrak{sl}(2, \mathbb{R})$ invariance subalgebra, but also see \cite{iibhora}, \cite{iibhorb}. So there may be a generalization of our results
to even-dimensional horizons. Alternatively, one may expect that there is an expression
relating the number of supersymmetries preserved by the even-dimensional horizons to the index
of a Dirac operator.

\acknowledgments

JG is supported by the STFC grant, ST/1004874/1.
GP is partially supported by the  STFC rolling grant ST/J002798/1.
The authors would like to thank J. Figueroa-O'Farrill, J.  Lotay and M. Singer for useful discussions.

\appendix

  \section{A Lichnerowicz Identity}

To prove the  Lichnerowicz identity of section 4, we use the spinor conventions of \cite{system11}, appendix A.2. In this conventions,  the Dirac spinors of $Spin(9,1)$ are identified  with $\Lambda^*(\mathbb{C}^5)$ and the Majorana spinors span a real 32-dimensional
 subspace after an appropriate reality condition is imposed. The Dirac spinors of $Spin(9)$ are identified with the subspace $\Lambda^*(\mathbb{C}^4)\subset \Lambda^*(\mathbb{C}^5)$. In particular,
  if $\mathbb{C}^5=\mathbb{C}<e_1,\dots e_4,e_5>$, then $\mathbb{C}^4=\mathbb{C}<e_1,\dots e_4>$. The Majorana spinors of $Spin(9)$ are those of $Spin(10,1)$ restricted on  $\Lambda^*(\mathbb{C}^4)$.
 From this, it is straightforward to identify the gamma matrices of $Spin(9)$ from those of $Spin(10,1)$ which have been given in \cite{system11}.

 As has been explained in section 4, the $Spin(9)$ invariant inner product $\langle \cdot , \cdot \rangle$ restricted on the Majorana representation is  positive definite and real, and so symmetric.   With respect to this,
 the skew-symmetric products $\Gamma^{[k]}$ of $k$ $Spin(9)$ gamma matrices are Hermitian for $k=0\, {\rm mod}\, 4$ and $k=1\, {\rm mod}\, 4$ while they are anti-Hermitian
 for $k=2\, {\rm mod}\, 4$ and $k=3\, {\rm mod}\, 4$. Using this, we have that
\begin{eqnarray}
\Psi^{(\pm)}{}_i^\dagger &=& \mp{1 \over 4} h_i -{1 \over 288} \Gamma_i{}^{\ell_1 \ell_2 \ell_3 \ell_4}
X_{\ell_1 \ell_2 \ell_3 \ell_4} -{1 \over 36} X_{i \ell_1 \ell_2 \ell_3} \Gamma^{\ell_1 \ell_2 \ell_3}
\mp{1 \over 24} \Gamma_i{}^{\ell_1 \ell_2} Y_{\ell_1 \ell_2} \mp{1 \over 6} Y_{ij} \Gamma^j~,
\nonumber \\
\Psi^{(\pm)\dagger} &=& \mp{1 \over 4} h_\ell \Gamma^\ell +{1 \over 96} X_{\ell_1 \ell_2 \ell_3 \ell_4}
\Gamma^{\ell_1 \ell_2 \ell_3 \ell_4} \mp{1 \over 8} Y_{\ell_1 \ell_2} \Gamma^{\ell_1 \ell_2}~,
\end{eqnarray}
where $\dagger$ is the adjoint with respect to the  $Spin(9)$-invariant inner product
$\langle \ , \ \rangle$.

Next let us turn to the computation of the RHS of  ({\ref{lichaux}}).
The term involving $\Psi^{(\pm)i \dagger} \Psi^{(\pm)}_i - \Psi^{(\pm)\dagger} \Psi^{(\pm)}$  can be expanded
out directly in terms quadratic in the fluxes $h, Y, X$. In
particular
\begin{eqnarray}
\label{quadpsi}
\Psi^{(\pm)i \dagger} \Psi^{(\pm)}_i &=& {1 \over 16} h^2 \pm{1 \over 576}
h_{\ell_1} X_{\ell_2 \ell_3 \ell_4 \ell_5} \Gamma^{\ell_1 \ell_2 \ell_3 \ell_4 \ell_5}
+{1 \over 12} (i_h Y)_\ell \Gamma^\ell
\nonumber \\
&-&{5 \over 27648} X_{\ell_1 \ell_2 \ell_3 \ell_4} X_{\ell_5 \ell_6 \ell_7 \ell_8}
\Gamma^{\ell_1 \ell_2 \ell_3 \ell_4 \ell_5 \ell_6 \ell_7 \ell_8}
-{1 \over 384} X^{mn}{}_{\ell_1 \ell_2} X_{mn \ell_3 \ell_4}
\Gamma^{\ell_1 \ell_2 \ell_3 \ell_4}
\nonumber \\
&+&{7 \over 1152} X_{\ell_1 \ell_2 \ell_3 \ell_4} X^{\ell_1 \ell_2 \ell_3 \ell_4}
+{1 \over 192} Y_{\ell_1 \ell_2} Y_{\ell_3 \ell_4} \Gamma^{\ell_1 \ell_2 \ell_3 \ell_4}
+{5 \over 96} Y^{mn} Y_{mn} \ .
\end{eqnarray}

The term in ({\ref{lichaux}}) involving $\Gamma^{ij} {\tilde{\nabla}}_i {\tilde{\nabla}}_j$  can be rewritten using
\begin{eqnarray}
\Gamma^{ij} {\tilde{\nabla}}_i {\tilde{\nabla}}_j \phi_\pm = -{1 \over 4}\tilde R \phi_\pm~,
\end{eqnarray}
where $\tilde R$ is the Ricci scalar of ${\cal{S}}$.  From the Einstein field equation ({\ref{ein1}}), one has
\begin{eqnarray}
\tilde R = - {\tilde{\nabla}}^i h_i +{1 \over 2} h^2 +{1 \over 4} Y_{mn} Y^{mn} +{1 \over 48} X_{\ell_1 \ell_2 \ell_3
\ell_4} X^{\ell_1 \ell_2 \ell_3 \ell_4}~.
\end{eqnarray}
It follows that
\begin{eqnarray}
 \int_{\cal{S}} \langle \phi_\pm , \Gamma^{ij} {\tilde{\nabla}}_i {\tilde{\nabla}}_j \phi_\pm \rangle
&=& \int_{\cal{S}} \langle \phi_\pm , \big(-{1 \over 8} h^2 -{1 \over 16} Y_{mn} Y^{mn}
-{1 \over 192} X_{\ell_1 \ell_2 \ell_3 \ell_4} X^{\ell_1 \ell_2 \ell_3 \ell_4} \big) \phi_\pm \rangle
\nonumber \\
&+& \int_{\cal{S}} {1 \over 4} {\tilde{\nabla}}^i h_i \langle \phi_\pm , \phi_\pm \rangle~.
\end{eqnarray}

To proceed with the  evaluation of ({\ref{lichaux}}), observe that
\begin{eqnarray}
\label{appaux1}
 \big( \Psi^{(\pm)i \dagger} &-& \Psi^{(\pm)i} \big) {\tilde{\nabla}}_i \phi_\pm
- \big( \Psi^{(\pm)\dagger} - \Psi^{(\pm)} \big) \Gamma^i {\tilde{\nabla}}_i \phi_\pm =
\big( -{1 \over 72} X_{\ell_1 \ell_2 \ell_3 \ell_4} \Gamma^{i \ell_1 \ell_2 \ell_3 \ell_4}
\mp{1 \over 6} Y^i{}_\ell \Gamma^\ell \big) {\tilde{\nabla}}_i \phi_\pm
\nonumber \\
&+& \big( {1 \over 72} X_{\ell_1 \ell_2 \ell_3 \ell_4} \Gamma^{\ell_1 \ell_2 \ell_3 \ell_4}
\pm{1 \over 6} Y_{\ell_1 \ell_2} \Gamma^{\ell_1 \ell_2} \big) \Gamma^i {\tilde{\nabla}}_i \phi_\pm~.
\nonumber \\
\end{eqnarray}
Using the fact that the Clifford algebra element of first term in the RHS of the above equation is hermitian, the Bianchi identity $dX=0$ and upon integrating by parts, one finds that
\begin{eqnarray}
 \int_{\cal{S}} \langle \phi_\pm , \big( \Psi^{(\pm)i \dagger} - \Psi^{(\pm)i} \big) {\tilde{\nabla}}_i \phi_\pm
- \big( \Psi^{(\pm)\dagger} - \Psi^{(\pm)} \big) \Gamma^i {\tilde{\nabla}}_i \phi_\pm \rangle
=\pm \int_{\cal{S}} \langle \phi, {1 \over 12} ({\tilde{\nabla}}^i Y_{i \ell}) \Gamma^\ell \phi \rangle
\nonumber \\
+  \int_{\cal{S}} \langle \phi , \big( {1 \over 72} X_{\ell_1 \ell_2 \ell_3 \ell_4}
\Gamma^{\ell_1 \ell_2 \ell_3 \ell_4} \pm{1 \over 6} Y_{\ell_1 \ell_2} \Gamma^{\ell_1 \ell_2}
\big) \Gamma^i {\tilde{\nabla}}_i \phi \rangle  \ .
\end{eqnarray}
The term involving ${\tilde{\nabla}}^i Y_{i \ell}$ is then further rewritten as a term quadratic in $X$ using the field equation
({\ref{geq2}}). Next, we rewrite the second line in terms of the Dirac operator $\Gamma^i {\tilde{\nabla}}_i \phi_\pm + \Psi^{(\pm)} \phi_\pm$,
with a compensating term $-\Psi^{(\pm)} \phi_\pm$ which gives a term quadratic in the fluxes $h, X, Y$, and
which can be expanded out straightforwardly.

Next, we find that
\begin{eqnarray}
\label{appaux2}
\big( \Gamma^i \Psi^{(\pm)} - \Psi^{(\pm)} \Gamma^i \big) {\tilde{\nabla}}_i \phi_\pm &=& \big( \mp{1 \over 2} h^i +{1 \over 48} \Gamma^{i \ell_1 \ell_2 \ell_3 \ell_4} X_{\ell_1 \ell_2 \ell_3 \ell_4} \pm{1 \over 2} Y^i{}_\ell \Gamma^\ell \big) {\tilde{\nabla}}_i \phi_\pm
\nonumber \\
&+& \big( \pm{1 \over 2} h_\ell \Gamma^\ell -{1 \over 48} X_{\ell_1 \ell_2 \ell_3 \ell_4}
\Gamma^{\ell_1 \ell_2 \ell_3 \ell_4} \big) \Gamma^i {\tilde{\nabla}}_i \phi_\pm~.
\end{eqnarray}
Similarly, using the hermiticity of the Clifford element in the first term in the RHS of the above equation, $dX=0$ and upon integrating by parts, one finds
\begin{eqnarray}
\int_{\cal{S}} \langle \phi_\pm , \big( \Gamma^i \Psi^{(\pm)} - \Psi^{(\pm)} \Gamma^i \big) {\tilde{\nabla}}_i \phi_\pm \rangle
&=& \int_{\cal{S}} \langle \phi_\pm , \mp{1 \over 4} ({\tilde{\nabla}}^i Y_{i \ell}) \Gamma^\ell \phi \rangle
\nonumber \\
&+&  \int_{\cal{S}} \langle \phi , \big( \pm{1 \over 2} h_\ell \Gamma^\ell
-{1 \over 48} X_{\ell_1 \ell_2 \ell_3 \ell_4} \Gamma^{\ell_1 \ell_2 \ell_3 \ell_4} \big)
\Gamma^i {\tilde{\nabla}}_i \phi \rangle
\nonumber \\
&\pm& {1 \over 4} \int_{\cal{S}} {\tilde{\nabla}}^i h_i \langle \phi_\pm , \phi_\pm \rangle
\ .
\nonumber \\
\end{eqnarray}
The term involving ${\tilde{\nabla}}^i Y_{i \ell}$ is then further rewritten as a term quadratic in $X$ using
({\ref{geq2}}). The second line is also
further rewritten  in terms of the Dirac operator $\Gamma^i {\tilde{\nabla}}_i \phi_\pm + \Psi^{(\pm)} \phi_\pm$,
with a compensating term involving $-\Psi^{(\pm)} \phi_\pm$ which gives a term quadratic in the fluxes $h, X, Y$.

Next note that
\begin{eqnarray}
(\Gamma^i {\tilde{\nabla}}_i \Psi^{(\pm)}- {\tilde{\nabla}}^i \Psi^{(\pm)}_i) \phi_\pm
&=& \bigg( \mp{1 \over 8} dh_{ij} \Gamma^{ij} +{1 \over 72} {\tilde{\nabla}}^i X_{i \ell_1 \ell_2 \ell_3}
\Gamma^{\ell_1 \ell_2 \ell_3}
\nonumber \\
&\pm&{1 \over 12} {\tilde{\nabla}}_{\ell_1} Y_{\ell_2 \ell_3}
\Gamma^{\ell_1 \ell_2 \ell_3} \pm{5 \over 12} {\tilde{\nabla}}^i Y_{i \ell} \Gamma^\ell \bigg) \phi_\pm~,
\nonumber \\
\end{eqnarray}
and so
\begin{eqnarray}
 \int_{\cal{S}} \langle \phi_\pm , (\Gamma^i {\tilde{\nabla}}_i \Psi^{(\pm)} - {\tilde{\nabla}}^i \Psi^{(\pm)}_i) \phi_\pm \rangle
= \int_{\cal{S}} \langle \phi_\pm , \pm{5 \over 12} {\tilde{\nabla}}^i Y_{i \ell} \Gamma^\ell  \phi_\pm \rangle \ ,
\end{eqnarray}
as the rest of the terms are anti-hermitian, and hence the associated form bilinears vanish.
The term involving ${\tilde{\nabla}}^i Y_{i \ell}$ is then again rewritten as a term quadratic in $X$ using
({\ref{geq2}}).

\section{Derivation of ({\ref{max2}})}

To derive ({\ref{max2}}), let us assume  that
$\phi_+$ satisfies the horizon Dirac equation ${\cal{D}}^{(+)} \phi_+ =0$.
Then
\begin{eqnarray}
\label{lapc1}
{\tilde{\nabla}}^i {\tilde{\nabla}}_i \parallel \phi_+\parallel^2 = 2  \langle \phi_+ , {\tilde{\nabla}}^i {\tilde{\nabla}}_i \phi_+ \rangle
+2 \langle {\tilde{\nabla}}_i \phi_+ {\tilde{\nabla}}^i \phi_+ \rangle~.
\end{eqnarray}
It will be useful to note the following identity
\begin{eqnarray}
{\tilde{\nabla}}^i {\tilde{\nabla}}_i \phi_+ &=& \Gamma^i {\tilde{\nabla}}_i (\Gamma^j {\tilde{\nabla}}_j \phi_+) - \Gamma^{ij} {\tilde{\nabla}}_i {\tilde{\nabla}}_j \phi_+
\nonumber \\
&=& \Gamma^i {\tilde{\nabla}}_i \bigg( \big({1 \over 4} h_j \Gamma^j -{1 \over 96} X_{\ell_1 \ell_2 \ell_3 \ell_4}
\Gamma^{\ell_1 \ell_2 \ell_3 \ell_4} -{1 \over 8} Y_{\ell_1 \ell_2} \Gamma^{\ell_1 \ell_2}  \big) \phi_+ \bigg)
+{1 \over 4} {\tilde{R}} \phi_+~.
\end{eqnarray}
This then implies that
\begin{eqnarray}
\label{apb1}
 \langle \phi_+, {\tilde{\nabla}}^i {\tilde{\nabla}}_i \phi_+ \rangle
&=& \langle \phi_+ , \big( {1 \over 8}h^2
+{1 \over 16} Y_{\ell_1 \ell_2} Y^{\ell_1 \ell_2}
+{1 \over 192} X_{\ell_1 \ell_2 \ell_3 \ell_4} -{1 \over 4} {\tilde{\nabla}}^i Y_{ij} \Gamma^j \big) \phi_+ \rangle
\nonumber \\
&+&   \langle \phi_+ , \Gamma^i \big({1 \over 4} h_j \Gamma^j
-{1 \over 96} X_{\ell_1 \ell_2 \ell_3 \ell_4} \Gamma^{\ell_1 \ell_2 \ell_3 \ell_4}
-{1 \over 8} Y_{\ell_1 \ell_2} \Gamma^{\ell_1 \ell_2}  \big) \phi_+ \rangle~,
\nonumber \\
\end{eqnarray}
where the trace of ({\ref{ein1}}) has been used.
Also, one has
\begin{eqnarray}
\label{apb2}
\langle {\tilde{\nabla}}^i \phi_+ , {\tilde{\nabla}}_i \phi_+ \rangle
&=& \langle {\tilde{\nabla}}^{(+)}{}^i \phi_+ , {\tilde{\nabla}}^{(+)}_i \phi_+ \rangle
-2  \langle \phi_+ , (\Psi^{(+)i})^\dagger {\tilde{\nabla}}_i \phi_+ \rangle
\nonumber \\
&-& \langle \phi_+ , \Psi^{(+)i \dagger} \Psi^{(+)}_i \phi_+ \rangle~.
\end{eqnarray}
Substituting ({\ref{apb1}}) and ({\ref{apb2}}) into ({\ref{lapc1}}), one obtains
\begin{eqnarray}
\label{apb3}
{\tilde{\nabla}}^i {\tilde{\nabla}}_i \langle \phi_+ , \phi_+ \rangle
&=&  \langle {\tilde{\nabla}}^{(+)}{}^i \phi_+ , {\tilde{\nabla}}^{(+)}_i \phi_+ \rangle
\nonumber \\
&+& \langle \phi_+ , \bigg({1 \over 4} h^2 +{1 \over 8} Y_{\ell_1 \ell_2} Y^{\ell_1 \ell_2}
+{1 \over 96} X_{\ell_1 \ell_2 \ell_3 \ell_4}X^{\ell_1 \ell_2 \ell_3 \ell_4}
\nonumber \\
&&-{1 \over 2} {\tilde{\nabla}}^i Y_{ij} \Gamma^j -2 \Psi^{(+)i \dagger} \Psi^{(+)}_i \bigg) \phi_+ \rangle
\nonumber \\
&+& 2  \ \langle \phi_+ \bigg( \Gamma^i \big( {1 \over 4} h_j \Gamma^j
-{1 \over 96} X_{\ell_1 \ell_2 \ell_3 \ell_4} \Gamma^{\ell_1 \ell_2 \ell_3 \ell_4}
\nonumber \\
 &&-{1 \over 8} Y_{\ell_1 \ell_2} \Gamma^{\ell_1 \ell_2} \big) -2 (\Psi^{(+)i})^\dagger \bigg) {\tilde{\nabla}}_i \phi_+ \rangle~.
\end{eqnarray}
To proceed, note that
\begin{eqnarray}
\Gamma^i \big( {1 \over 4} h_j \Gamma^j
-{1 \over 96} X_{\ell_1 \ell_2 \ell_3 \ell_4} \Gamma^{\ell_1 \ell_2 \ell_3 \ell_4}
-{1 \over 8} Y_{\ell_1 \ell_2} \Gamma^{\ell_1 \ell_2} \big) -2 (\Psi^{(+)i})^\dagger
=h^i
\nonumber \\
\qquad \qquad + \bigg(-{1 \over 4} h_j \Gamma^j -{1 \over 288} X_{\ell_1 \ell_2 \ell_3 \ell_4}
\Gamma^{\ell_1 \ell_2 \ell_3 \ell_4} -{1 \over 24} Y_{\ell_1 \ell_2} \Gamma^{\ell_1 \ell_2}
\bigg)\Gamma^i~.
\end{eqnarray}
Substitute this expression into ({\ref{apb3}}), and use the Dirac equation
${\cal{D}}^{(+)} \phi_+ =0$ to eliminate the $\Gamma^i {\tilde{\nabla}}_i \phi_+$ term in
favour of terms algebraic in the fluxes. On expanding out the resulting expression
and making use of ({\ref{quadpsi}}) and ({\ref{ydiv}}), one then obtains ({\ref{max2}}).

\end{document}